\documentclass[final,5p,times,twocolumn,number]{elsarticle}

\usepackage{amssymb}
\usepackage{lipsum}
\usepackage{xcolor}
\usepackage{amsmath}

\biboptions{sort&compress}

\journal{Physics Letters B}

\begin{document}

\begin{frontmatter}
\title{Effective Phantom Divide Crossing with Standard and Negative Quintessence}

\author{Adri\`a G\'omez-Valent}
\ead{agomezvalent@icc.ub.edu}
\author{Alex Gonz\'alez-Fuentes}
\ead{agonzalezfuentes@icc.ub.edu}
\address{Departament de F\'isica Qu\`antica i Astrof\'isica, and Institut
de Ciències del Cosmos, \\Universitat de Barcelona, Av.\ Diagonal 647, E-08028 Barcelona,
Catalonia, Spain}

\begin{abstract}
Cosmic microwave background data from the {\it Planck} satellite, combined with baryon acoustic oscillation measurements from the Dark Energy Spectroscopic Instrument and Type Ia supernovae from various samples, provide hints of dynamical dark energy (DE). These results indicate a peak in the DE density around $z\sim 0.4-0.5$, with the highest significance observed when using the supernovae from the Dark Energy Survey. In this {\it Letter}, we show that this peak does not necessarily imply a true crossing of the phantom divide if the measured effective DE is not a single component, but a combination of standard and negative quintessence. The latter is characterized by negative energy density and positive pressure, both decreasing in absolute value and tending to 0 in the future. For appropriate values of the parameters, negative quintessence is relevant at intermediate redshifts and becomes subdominant in front of standard quintessence around $z\sim 0.4-0.5$, giving rise to the aforementioned peak in the DE density. We find that our model is preferred over $\Lambda$CDM at a $3.26\sigma$ CL, which is comparable to the level of exclusion found with the Chevallier-Polarski-Linder parametrization. Our analysis leaves open the possibility of negative quintessence and other exotic fields existing in the low-energy universe, potentially playing a significant role in cosmic dynamics.
\end{abstract}

\end{frontmatter}

\section{Introduction}
\label{introduction}

 Several studies over the past decade have reported indications of dynamical dark energy (DE) in the context of the cosmological tensions \cite{Sahni:2014ooa,Salvatelli:2014zta,Sola:2015wwa,Sola:2016jky,SolaPeracaula:2016qlq,Zhao:2017cud,SolaPeracaula:2017esw,Sola:2017znb,SolaPeracaula:2018wwm}, and recent observations from Type Ia supernovae (SNIa), baryon acoustic oscillations (BAO), and the cosmic microwave background (CMB) suggest a non-trivial evolution of the effective DE density, with a transition in the DE behavior occurring around $z_{\rm cross} \sim 0.4$–$0.5$, from phantom-like at earlier times ($z > z_{\rm cross}$) to quintessence-like at lower redshifts ($z < z_{\rm cross}$) \cite{DESI:2025zgx,DESI:2025fii,Gonzalez-Fuentes:2025lei,Li:2025ops}. Evidence for this crossing of the phantom divide is statistically significant, with confidence levels ranging from 96.21\% to 99.97\%, depending on the SNIa dataset used \cite{Gonzalez-Fuentes:2025lei}; see also \cite{Lu:2025gki,Keeley:2025rlg,Ozulker:2025ehg,Silva:2025twg}.

Constructing theoretical models that reproduce this behavior is an important challenge. It is well known that a single minimally coupled scalar field cannot account for such a crossing, see, e.g., \cite{Fang:2008sn,Cai:2009zp,Cai:2025mas}. Other scenarios, like those involving non-minimally coupled scalar fields \cite{Ye:2024ywg,Ye:2024zpk,Tiwari:2024gzo,Wolf:2025jed}, interactions in the dark sector \cite{Chakraborty:2025syu,Khoury:2025txd,Guedezounme:2025wav,Chen:2025ywv},  braneworld dark energy \cite{Mishra:2025goj}, $F(R,T,Q)$ and other modifications of gravity \cite{Odintsov:2024woi,Yang:2025mws,Nojiri:2025low,Yao:2025wlx,Tsujikawa:2025wca}, exotic dark matter \cite{Chen:2025wwn,Giani:2025hhs,Braglia:2025gdo} or deviations from the cosmological principle \cite{Camarena:2025upt} might be viable possibilities. In this paper, we propose an alternative framework with two scalar fields minimally coupled to gravity and demonstrate that our model is perfectly capable of explaining the latest CMB, BAO and SNIa observations. 

\section{The model}

We consider the following action,

\begin{equation}\label{eq:action}
S \supset  -\int d^4x\sqrt{-g}\sum_{I=1,2}\left[\frac{\alpha_I}{2}\partial_\mu\phi_I\partial^\mu\phi_I+V_I(\phi_I)\right]\,, 
\end{equation}
with $V_I(\phi_I)$ the scalar field potentials and $\alpha_I$ constants equal to $+1$ or $-1$, depending on the specific model we want to focus on. The subscript $I=1,2$ labels the two scalar fields. Notice that we employ natural units and only use Einstein's summation convention for the Greek indices. The total action $S$ also includes the Einstein-Hilbert term, the standard model of particle physics and some potential extensions of it accounting, among other things, for the neutrino masses and the dark matter sector.  

The modified Klein-Gordon (KG) equation that rules the scalar field dynamics in a Friedmann-Lema\^itre-Robertson-Walker universe takes the form,

\begin{equation}\label{eq:KG}
\ddot{\phi}_I+3H\dot{\phi}_I+\frac{1}{\alpha_I}\frac{\partial V_I}{\partial\phi_I} = 0\,,
\end{equation}
where the dots denote derivatives with respect to cosmic time. The energy density and pressure of the individual scalar fields read, respectively,

\begin{equation}\label{eq:densitypressure}
    \rho_I = \alpha_I\frac{\dot{\phi}^2_I}{2}+V_I(\phi_I)\quad ;\quad p_I = \alpha_I\frac{\dot{\phi}^2_I}{2}-V_I(\phi_I)\,,
\end{equation}
while the time derivative of the energy densities is given by 
\begin{equation}\label{eq:der}
\dot{\rho}_I=-3H\alpha_I\dot{\phi}_I^2\,,
\end{equation}
indicating that the sign of the parameter $\alpha_I$ controls whether $\rho_I$ grows or gets diluted throughout the expansion, since $H\dot{\phi}^2_I\geq 0$. This holds regardless of the potential's sign or shape. Therefore, given the dark energy phenomenology required to accommodate the current CMB, BAO, and SNIa data, and in line with what was discussed in the Introduction, it is clear that we must at least require $\alpha_1=-\alpha_2$ in order to obtain a period during which the dark energy density increases with the expansion followed by a period in which it decreases. The kinetic terms of the two scalar fields must have then opposite signs. There is no way out. 

However, this condition is necessary, but certainly not sufficient to explain the data. Let us choose the field $\phi_2$ to be the one with the non-standard kinetic term, i.e., we set $\alpha_2=-1$ and $\alpha_1=+1$. Obviously, the potentials $V_I(\phi_I)$ will play an important role as well, since they control the dynamics of the fields through Eq. \eqref{eq:KG}. For instance, if this dynamics is activated too late or too soon in the cosmic history, or if the dynamics of $\phi_2$ is activated after that of $\phi_1$ it is clear that the model will not be capable of explaining the observations. For simplicity, let us consider quadratic potentials, 

\begin{equation}
V_I(\phi_I) = V_{0,I}+\frac{\epsilon_I}{2}m_I^2\phi_I^2\,,
\end{equation}
with $V_{0,I}$ constants and  $m_I$ the masses of the fields, which control the curvature of the potential and, hence, the moment in cosmic history at which the fields become dynamical. The constant $\epsilon_I$ is $+1$ or $-1$ for concave and convex potentials, respectively. 

At sufficiently large redshifts, the scalar field  potentials play no important role because cosmic friction dominates over the potential term in the KG equation \eqref{eq:KG}, so $\dot{\phi}_I\sim 0$. At those epochs, $\phi_1$ and $\phi_2$ are frozen and these two components essentially behave jointly as a  cosmological constant with value 

\begin{equation}\label{eq:Vini}
V_{\rm ini} =\sum_{I=1,2}\left[ V_{0,I}+\frac{\epsilon_I}{2}m_I^2\phi_{I,\rm ini}^2\right]\,.
\end{equation}
The scalar field $\phi_I$ starts to ``move'' when $H\simeq m_I$, exhibiting the typical thawing behavior \cite{Caldwell:2005tm} when $\alpha_I=1$. For our model to produce an effective crossing of the phantom divide from phantom (at large $z$) to quintessence (at small $z$), we obviously need to activate first the dynamics of $\phi_2$. Therefore, we need $m_2>m_1$. When this happens, $\phi_1$ remains frozen and $\phi_2$ starts to evolve. In this period, the DE component behaves effectively as phantom DE, since its energy density grows with time. When $H\simeq m_1$, $\phi_1$ also acquires dynamics and, at some moment, and according to  Eq. \eqref{eq:der},  $|\dot{\phi}_2|$ has to decrease below $|\dot{\phi}_1|$ in order to give rise to an effective quintessence era. The equality $\dot{\phi}_1^2=\dot{\phi}_2^2$ happens at the transition redshift $z_{\rm cross}$ -- when the effective phantom divide crossing occurs --, but only if the potential of $\phi_2$ is convex, i.e., if $\epsilon_2=-1$. There is no strong requirement on the sign of $\epsilon_1$, though. It can be both positive or negative. To summarize, these are some necessary conditions that should be fulfilled for a successful phenomenological performance of the model, in the light of current data: (i) $\alpha_1=-\alpha_2$; and if we choose $\alpha_2=-1$, (ii) $m_2>m_1$; and (iii) $\epsilon_2=-1$. Thus, we need to consider,

\begin{figure}[t!]
    \centering    \includegraphics[width=0.85\linewidth]{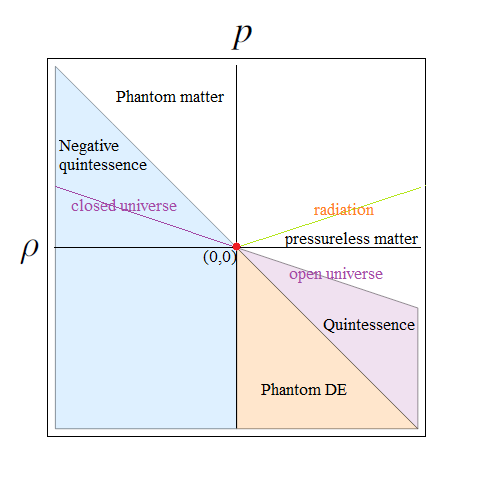}
    \caption{EoS diagram. The pink and orange regions correspond to the SQ ($-1\leq w\leq-1/3$) and phantom ($w\leq-1$) DE regimes, respectively. In the white region the strong energy condition is fulfilled, i.e., $\rho+3p\geq 0$ and $\rho+p\geq 0$. In particular, non-relativistic matter ($w=0$) and radiation ($w=1/3$) lie within this region, satisfying $p,\rho\geq 0$. Phantom matter ($w\leq -1$) as well, with $\rho<0$ and $p>0$  \cite{Grande:2006nn,Mavromatos:2021urx,Gomez-Valent:2024tdb,Gomez-Valent:2024ejh}. The blue region is the one for species with negative energy density and decreasing absolute value. Within that region, the triangular area bounded by the ``closed universe'' line and the edge of the ``phantom matter'' region represents the domain of the NQ scenario.}
    \label{fig:EoSdiag}
\end{figure}

\begin{equation}
V_1(\phi_1) = V_{0,1}\pm\frac{m_1^2}{2}\phi_1^2\qquad ;\qquad V_2(\phi_2) = V_{0,2}-\frac{m_2^2}{2}\phi_2^2\,.
\end{equation}
The data is not sensitive to $V_{0,1}$ and $V_{0,2}$ individually, but to their sum, i.e., to $V_0\equiv V_{0,1}+V_{0,2}$. Therefore, we have  the freedom to use different setups for the scalar field potentials, with each of these setups associated to a different splitting of $V_0$. For instance, if we set $V_{0,2}$ to zero and for typical values of the constant $V_0$ able to explain the current acceleration of the universe, $\phi_1$ is as a standard quintessence (SQ) field, while $\phi_2$ behaves in a quite exotic way, which is not as standard quintessence nor phantom DE do. Its energy density is negative and grows with the expansion -- decreases in absolute value -- due to the fact that $\alpha_2<0$, and its pressure is positive and decreases with cosmic time. The scalar field $\phi_2$ evolves towards the maximum of $V_2(\phi_2)$. We coin this entity negative quintessence (NQ), as in \cite{Gonzalez-Fuentes:2025lei}, since it has a negative potential and negative (non-standard) kinetic term. In the equation-of-state (EoS) diagram of Fig. \ref{fig:EoSdiag}, NQ lies in the region between the closed universe ($w=-1/3$) and phantom ($w=-1$) lines, with negative energy density and positive pressure. We call this model involving SQ and NQ SQ+NQ, for short. This is the model we will focus on in this {\it Letter}. Lagrangians of this sort were previously studied in \cite{Kaplan:2005rr}. 

NQ might be affected by the same issues as phantom DE, owing to the presence of instabilities associated with rapid vacuum decay into ghosts and gravitons. This is a contrived discussion in the absence of a complete theory of quantum gravity. We treat the model as a low-energy effective theory, in the hope that either higher-order derivative terms in the action or the presence of an energy cutoff associated with Lorentz invariance will render the theory stable at the quantum level (see \cite{Carroll:2003st,Cline:2003gs,Kaplan:2005rr,Emparan:2005gg,Garriga:2012pk,Gross:2020tph,Amendola:2015ksp} for detailed discussions). Another possibility is that gravity cannot be quantized and, consequently, gravitons do not exist, thereby rendering vacuum decay into ghosts impossible.

At this point, it is important to remark that other splitting options of $V_0$ exist. We could set $V_{0,1}=0$, instead of $V_{0,2}=0$. In this case, $\phi_2$ would be an ordinary phantom DE field \cite{Caldwell:1999ew} and $\phi_1$ a SQ field if $\epsilon_1=+1$, or a phantom matter field if $\epsilon_1=-1$. Phantom matter also has $\rho<0$ and $p>0$ (as NQ), but their absolute values grow with the expansion, instead of decreasing, since $w<-1$ \cite{Grande:2006nn,Mavromatos:2021urx,Gomez-Valent:2024tdb,Gomez-Valent:2024ejh}, cf. again Fig. \ref{fig:EoSdiag}. Other hybrid options are possible, of course, if one considers the two constants $V_{0,1},V_{0,2}\ne 0$ and plays with their signs. While the physical interpretation of the entities constituting the composite dark energy sector differs among these scenarios -- since the corresponding species occupy completely different regions of the EoS diagram -- they lead to a very similar background phenomenology, and represent viable alternatives to the SQ+NQ model. They mimic quintom-B models \cite{Cai:2009zp,Cai:2025mas},  allowing for the crossing of the phantom divide in the desired direction, i.e., from effective phantom at high redshift to quintessence at low redshift \footnote{This is opposed to quintom-A models \cite{Feng:2004ad,Chimento:2008ws,Cai:2009zp,Cai:2025mas}, which lead to a crossing of the phantom divide from quintessence to phantom.}.

Let us come back now to the SQ+NQ model. At the perturbative level, neither SQ nor NQ exhibit tachyonic nor gradient instabilities. This is easy to understand by looking at the equation for the scalar field perturbation $\delta\phi_I$, which in the synchronous gauge reads, 

\begin{equation}
\delta\ddot{\phi}_I+3H\delta\dot{\phi}_I+\left(c_s^2k^2+m_I^2\right)\delta\phi_I+\frac{\dot{h}\dot{\phi_I}}{2}=0\,,
\end{equation}
with $h$ the trace of the metric perturbation. There is no exponential growth of $\delta\phi_I$ at any scale, since both $m_I^2>0$ and $c_s^2=1>0$. These instabilities are not dramatic in general if they develop in a timescale $\gtrsim \mathcal{O}(H^{-1})$, but in our model they are simply non-existent. Dark energy does not cluster efficiently in our model, and its impact on the large-scale structure of the universe is essentially mediated through changes in the background dynamics. 

Some comments are also in order regarding the form of the action assumed for the SQ+NQ model. The presence of a linear interaction term between $\phi_1$ and $\phi_2$, i.e. $\lambda \phi_1 \phi_2$, would not affect the dynamics of the model, since it can always be eliminated by diagonalizing the system, thereby recovering an action equivalent to the one considered in this work. On the other hand, we do not include cubic or quartic self-interaction terms, or interactions between the two scalar fields beyond linear order. Although such extensions could, in principle, be introduced, they would add additional parameters to the model that, as shown below, are not required for an accurate description of the data. The absence of interaction terms is consistent with our framework if the corresponding dimensionless couplings $\beta_i\ll(H_0/M_{\rm pl})^2$, with $M_{\rm pl}$ the Planck mass. This requires the usual degree of fine-tuning that is already present in standard DE models. Kinetic mixing terms of the type $\partial_\mu\phi_1\partial^\mu\phi_2$ could play a role in the late universe, but again they have been neglected in our analysis for the sake of simplicity.

Let us recapitulate. At the background level, the dynamics of the SQ+NQ model works as follows. Initially, the fields are frozen, so they contribute as a cosmological constant through the following constant potential,

\begin{equation}\label{eq:Vini}
V_{\rm ini} =V_0+\frac{1}{2}\left(m_1^2\phi_{1,\rm ini}^2-m_2^2\phi_{2,\rm ini}^2\right)\,.
\end{equation}
When $H\simeq m_2$, $\phi_2$ starts to climb up $V_2(\phi_2)$. Its energy density evolves to less negative values and its pressure to less positive values. As a consequence, the total energy density of the DE sector increases during this epoch. Finally, if $m_1>H_0$, SQ starts to roll down $V_1(\phi_1)$ in the past -- after the activation of NQ -- and eventually, for appropriate initial conditions and masses, the energy density of NQ becomes derisory in front of that of SQ. From that moment until the present, the total DE behaves as quintessence. This transition occurs at redshift $z_{\rm cross}$, defined by the condition $\dot{\phi}_2^2(z_{\rm cross}) = \dot{\phi}_1^2(z_{\rm cross})$. In the future, $\phi_1$ will oscillate around the minimum of $V_1(\phi_1)$ and $\phi_2$ around the maximum of $V_2(\phi_2)$, leading to dilution laws of the form $\rho_I(a)\sim a^{-3}$ -- negative in the case of NQ -- and zero pressure.

In the matter-dominated epoch onward, the background dynamics are ruled by the following system of differential equations formed by the Friedmann and the KG equations, 

\begin{equation}
   E^2(a) = \frac{\Omega_{\rm m}^0 a^{-3}+\Omega_\Lambda^0+\frac{1}{6}\left(M_1^2\varphi_1^2-M_2^2\varphi_2^2\right)}{1+\frac{a^2}{6}\left[(\varphi_2^\prime)^2-(\varphi_1^\prime)^2\right]} \,,
\end{equation}

\begin{equation}
\varphi_I^{\prime\prime}+\varphi_I^\prime\left[\frac{4}{a}-\frac{3\Omega_{\rm m}^0a^{-4}}{2E^2(a)}-\frac{a}{2}\left((\varphi_1^\prime)^2-(\varphi_2^\prime)^2\right)\right]+\frac{M_I^2\varphi_I}{a^2E^2(a)}=0\,,
\end{equation}
with $\Omega_{\rm m}^0=\rho_{\rm m}^0/\rho_c^0$ and  $\Omega_\Lambda^0\equiv V_0/\rho_c^0$, and $\rho_c^0$ the current critical energy density in the universe. The primes denote derivatives with respect to the scale factor. For convenience, we use the following dimensionless quantities, $E\equiv H/H_0$, $M_I\equiv m_I/H_0$ and $\varphi_I \equiv (8\pi G)^{1/2} \phi_I$. The contribution of radiation can be added trivially. In practice, for the mass values of interest, the model reduces to $\Lambda$CDM during radiation domination, since $V_{\rm ini} \ll \rho_r$ (cf. Eq.~\ref{eq:Vini}), and thus has no impact on the cosmic dynamics.

The model has four additional parameters compared to $\Lambda$CDM, namely, $M_1$, $M_2$, and the two initial field values, $\varphi_{1,\rm ini}$ and $\varphi_{2,\rm ini}$. It reduces to the standard model if $M_1,M_2\ll H_0$ or if the initial values of $\varphi_1$ and $\varphi_2$ sit already on the minimum of $V_{1}$ and the maximum of $V_2$, respectively. The initial values of the field derivatives, $\dot{\varphi}_{1,\rm ini}$ and $\dot{\varphi}_{2,\rm ini}$, are naturally set to zero, since this is an attractor in the radiation-dominated epoch, where the initial conditions are imposed; and $\Omega_\Lambda^0$ is fixed through a shooting method to fulfill the consistency condition $E(a=1)=1$.

\section{Data and methodology}

We employ state-of-the-art data from CMB, BAO and SNIa to constrain our model. In particular, we make use of compressed CMB information  through the same correlated Gaussian prior on $(\theta_*,\omega_{\rm b},\omega_{\rm bc})_{\rm CMB}$ as in \cite{DESI:2025zgx}, which is based on {\it Planck}'s \texttt{CamSpec} CMB likelihood \cite{Lemos:2023xhs}. This is sufficient, since our model only introduces new physics in the late universe. We also use the BAO data from DESI Data Release 2 (DR2) \cite{DESI:2025zgx} and the SNIa from DES-Y5 \cite{DES:2024hip,DES:2024jxu}. We restrict our exploration of the parameter space to the region where $\varphi_I$ exhibit no oscillations at present by using the prior $0\leq M_1<M_2\leq10$.

Our model  is implemented in a modified version of \texttt{CLASS} \cite{lesgourgues2011cosmiclinearanisotropysolving, Diego_Blas_2011}, which allows us to compute all the needed background observables. The exploration of the parameter space and the extraction of the Monte Carlo Markov chains is carried out with \texttt{Cobaya} \cite{Torrado:2020dgo}, and the resulting chains are analyzed with \texttt{GetDist} \cite{Lewis:2019xzd}.

We employ two complementary approaches to assess the relative fitting performance of our model with respect to $\Lambda$CDM. First, we use the likelihood-ratio test \cite{LRtest,Wilk1938} to estimate the exclusion level of $\Lambda$CDM relative to the alternative model in terms of the number of sigmas, following, e.g., \cite{DESI:2025zgx,Gonzalez-Fuentes:2025lei}. We denote this quantity as $E_{\Lambda{\rm CDM}}$. In addition, we use the Akaike information criterion (AIC) \cite{Akaike}, which is defined as

\begin{equation}\label{eq:AIC}
{\rm AIC} = \chi^2_{\rm min} + 2N\,,
\end{equation}
with $\chi^2_\mathrm{min}$ the minimum value of $\chi^2$ and $N$ the total number of parameters of the model. The AIC is a mathematical implementation of Occam's razor. It decreases with lower values of $\chi^2_{\rm min}$ and increases with the number of model parameters, thereby penalizing greater model complexity. We define the difference $\Delta {\rm AIC}_j \equiv {\rm AIC}_{\Lambda{\rm CDM}}-{\rm AIC}_j$ between the $\Lambda$CDM value of AIC and that of a model $j$. According to Jeffreys' scale, if this difference is larger than $5$ it indicates a strong evidence for the model with the lowest values of AIC; and very strong evidence if the difference is larger than $10$\footnote{Other information criteria, such as the Bayesian \cite{BIC} or deviance \cite{DIC} information criteria, do not perform properly in this analysis -- either because they are overly conservative, to the point of conflicting with the more robust likelihood-ratio test, with penalization factors more than three times the one employed in AIC (this is the case of BIC), or because they fail in the presence of significant volume effects (as in DIC). As we will see in Sec. \ref{sec:results}, volume effects are non-negligible in the SQ+NQ model.}.

\begin{figure*}[t!]
    \centering    \includegraphics[width=\linewidth]{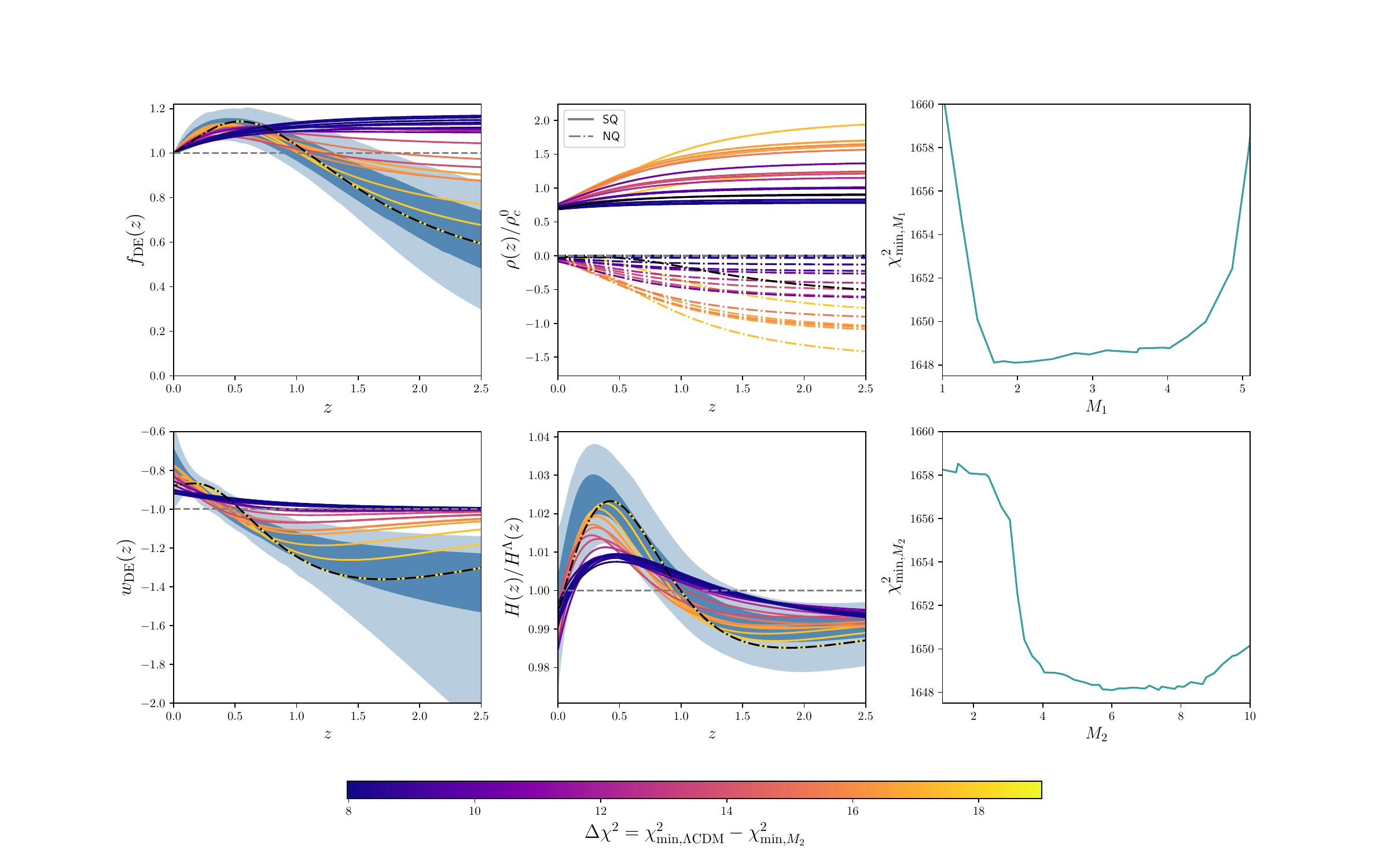}
    \caption{{\it Right plots:} Profile distributions for $M_1$ and $M_2$. A wide range of masses lead to low values of $\chi^2$; {\it Left and central plots:} In addition to the reconstructions obtained with Planck+DESI+DES-Y5 \cite{Gonzalez-Fuentes:2025lei} for the relevant background quantities (shown in light blue), we display several curves associated to SQ+NQ models with varying ability to describe the data. We use models corresponding to different points on the profile likelihood of M2, with the best-fit model shown as a black dot-dashed line. For the $\Lambda$CDM $H^\Lambda(z)$ we set $\Omega_{\rm m}^0=0.315$ and $H_0=67.26$ km/s/Mpc \cite{Rosenberg:2022sdy}. In the central upper plot, we show the SQ and NQ densities that contribute to the total DE density in solid and dashed curves, respectively.}
    \label{fig:example}
\end{figure*}

\section{Results}\label{sec:results}

Our fitting results are displayed in Table \ref{tab:fitting_tab}. We compare the performance of our model with those of $\Lambda$CDM and the Chevallier-Polarski-Linder (CPL) parametrization, in which DE is modeled as a perfect fluid with EoS $w(a)=w_0+w_a(1-a)$ \cite{Chevallier:2000qy,Linder:2002et}. We find that the CPL model yields a decrease in $\chi^2_\mathrm{min}$ of approximately 19 units relative to $\Lambda$CDM, corresponding to reductions of about 15 units in the AIC. This suggests a very strong evidence for the CPL. The likelihood-ratio test also indicates a preference for the CPL model at $3.93\sigma$ CL, which is in perfect agreement with previous works, see, e.g., \cite{DESI:2025zgx,Gonzalez-Fuentes:2025lei,Giare:2025pzu}, and also with AIC. The $\chi^2_\mathrm{min}$ decreases at a similar level -- by $\sim$18 units compared to $\Lambda$CDM -- when moving to the SQ+NQ model. Taking into account that the latter has 4 more parameters than the $\Lambda$CDM, the AIC in this case is $\sim10$ units smaller than in the standard model. This translates into a preference of SQ+NQ over $\Lambda$CDM at $3.26\sigma$ CL.  

We have checked that our model suffers from non-negligible volume effects. The one-dimensional profile likelihoods for $M_1$ and $M_2$ shown in the right plots of Fig. \ref{fig:example} -- obtained following the method of Ref. \cite{Gomez-Valent:2022hkb} -- demonstrate that there is a wide range of masses -- namely, $2\lesssim M_1\lesssim 4$, $4\lesssim M_2\lesssim 9$ -- that lead to low values of $\chi^2$. In any case, these marginalization effects do not bias our conclusions, as the latter are solely based on the $\chi^2_{\rm min}$ values of each model.

The left and central plots of Fig. \ref{fig:example} show that our model is perfectly able to produce shapes of $H(z)$ and the normalized DE energy density $f_{\rm DE}(z)=\rho_{\rm DE}(z)/\rho_{\rm DE}^0$ in the $1\sigma$ region of the model-agnostic reconstructions obtained using CMB+BAO+SNIa \cite{Gonzalez-Fuentes:2025lei}. Notice that neither of the two components forming the composite DE sector crosses the phantom divide, cf. the central upper plot. However, the effective EoS parameter, defined as $w_{\rm DE}(z)=(p_1(z)+p_2(z))/(\rho_1(z)+\rho_2(z))$ does exhibit a transition from phantom -- despite there is no phantom DE -- to quintessence. Fig. \ref{fig:example} also demonstrates that thawing quintessence can fit the data better than $\Lambda$CDM but falls short in front of models with crossing, a conclusion reached also by other authors \cite{Wolf:2024eph,Gialamas:2025pwv,Cline:2025sbt,Ozulker:2025ehg}.

\begin{table}
\begin{center}
\begin{tabular}{|c|ccc|}
 \hline
 Parameter & $\Lambda$CDM & CPL & SQ+NQ \\ \hline
$\Omega_\mathrm{m}^0$ & $0.303\pm0.004$ & $0.319\pm0.006$ & $0.315\pm0.006$ \\
$H_0$ & $68.10$& $66.70\pm0.58$ &  $67.04\pm 0.60$ \\
$M_1$ & $-$ & $-$ & $2.5^{+0.4}_{-0.9}$ \\
$M_2$ & $-$ & $-$ & $6.3^{+1.7}_{-2.0}$ \\
$\varphi_{1,{\rm ini}}$ & $-$ & $-$ & $0.66^{+0.18}_{-0.23}$ \\
$\varphi_{2,{\rm ini}}$ & $-$ & $-$ &  $0.35^{+0.06}_{-0.12}$\\
$w_0$ & $-$ & $-0.756\pm0.059$ & $-$ \\  
$w_a$ & $-$ &  $-0.84\pm0.24$ & $-$  \\ \hline 
$\chi^2_\mathrm{min}$ &  $1666.36$&  $1647.65$ &  $1648.11$ \\
$\Delta$AIC & $-$ & $14.71$ & $10.25$ \\
$E_{\Lambda{\rm CDM}}$ & $-$ &  $3.93\sigma$ &  $3.26\sigma$\\\hline
\end{tabular}
\caption{Mean values and uncertainties at 68\% CL obtained using Planck+DESI+DES-Y5 for the $\Lambda$CDM, CPL and SQ+NQ, from the corresponding one-dimensional posterior distributions. We also display the minimum values of $\chi^2$, the differences between the Akaike information criteria of $\Lambda$CDM and the other models --- cf. Eq. \eqref{eq:AIC} --, and the statistical exclusion level of $\Lambda$CDM inferred from the likelihood-ratio test.}\label{tab:fitting_tab}
\end{center}
\end{table}

We remark that our results are highly non-trivial. The phenomenology required to fit the data cannot be reproduced by two perfect fluids with constant EoS parameters plus a cosmological constant, even if one of the fluids is allowed to have negative energy density and positive pressure, as NQ. Let us show that explicitly. In such a scenario, the normalized DE density reads,

\begin{equation}
f_{\rm DE}(z) = \frac{1-\Omega_{\rm m}^0+\Omega_1^0[(1+z)^{\xi_1}-1]+\Omega_2^0[(1+z)^{\xi_2}-1]}{1-\Omega_{\rm m}^0}\,,
\end{equation}
with $\xi_i\equiv3(1+w_i)$ for $i=1,2$. We want this function to satisfy, at least, these three conditions: (i) $f(z=0.4)\sim 1.1$; (ii) $df/dz=0$ at $z\sim 0.4$; and $f(z\sim 1.1)=0$. All of them are fulfilled in good approximation by our reconstruction \cite{Gonzalez-Fuentes:2025lei}, cf. Fig. \ref{fig:example}. Using the conditions (i)+(ii) and (ii)+(iii) we obtain, respectively,

\begin{equation}
\Omega_1^0(\xi_1,\xi_2)=\frac{0.1(1-\Omega_{\rm m}^0)}{1.4^{\xi_1}-1-\frac{\xi_1}{\xi_2}1.4^{\xi_1-\xi_2}(1.4^{\xi_2}-1)}\,.
\end{equation}

\begin{equation}
\Omega_1^0(\xi_1,\xi_2)=\frac{\Omega_{\rm m}^0-1}{2.1^{\xi_1}-1-\frac{\xi_1}{\xi_2}1.4^{\xi_1-\xi_2}(2.1^{\xi_2}-1)}\,,
\end{equation}
By equating these two expressions we could find in principle an equation that relates $\xi_1$ and $\xi_2$. However, they describe two-dimensional surfaces that do not intersect each other, so there is no point $(\xi_1,\xi_2)$ that fulfills the required conditions. This is telling us that, besides the presence of NQ and its freezing behavior at low redshift, the thawing nature of the SQ field is also important for the success of the SQ+NQ model.

\section{Conclusions}

The persistent hints of dynamical dark energy offer a valuable opportunity in the quest to understand its fundamental nature. With this in mind, it is essential to distinguish which features are genuinely supported by the data, rather than being spurious artifacts of specific models or parametrizations. The CPL parametrization provides a great fit to the current CMB+BAO+SNIa data, and it suggests a crossing of the phantom divide, which has caused much concern. On the other hand, it is well known that if DE consists of a single scalar field slowly rolling down its potential its EoS cannot cross $w=-1$, although a fit through the lens of CPL could lead us to the opposite conclusion \cite{Wolf:2023uno}. To more precisely assess the degree of evidence for a crossing of the phantom divide, we applied in \cite{Gonzalez-Fuentes:2025lei} a model-agnostic reconstruction of the DE EoS and its energy density and quantified the probability of crossing to be of 96.21\% to 99.97\%, see also \cite{Lu:2025gki,Keeley:2025rlg,Ozulker:2025ehg, Silva:2025twg}. In view of these results, it is worthwhile to investigate whether the peak observed in $f_{\rm DE}$ at $z\sim 0.4-0.5$ can be realized within a model derived from an action-based formulation.  

In this {\it Letter}, we show that this phenomenology can be produced without a true crossing of the phantom divide, considering only two minimally coupled scalar fields\footnote{In Ref. \cite{Goh:2025upc}, a similar model was studied. Their preprint appeared on arXiv approximately one month and a half after ours.}. The effective DE resulting from the combination of standard and negative quintessence can replicate the required shape of the background quantities of interest. NQ has negative energy density and positive pressure, both of which decrease in absolute value with the expansion. The two fields are characterized by their mass and initial field value (besides one positive cosmological constant). We compare the performance of this model with $\Lambda$CDM and the CPL parametrization. Our model outperforms $\Lambda$CDM -- by $\sim 18$ units of $\chi^2_{\rm min}$ -- and describes the CMB+BAO+SNIa data at a similar level to CPL -- the latter is only moderately preferred over SQ+NQ. The standard model is excluded at  $3.93\sigma$ for CPL and $3.26\sigma$ for SQ+NQ. The slight decrease of significance in our model is due to the introduction of two additional parameters. However, our model arises from an action and, in principle, it could be regarded as a low-energy limit of a more fundamental (UV-complete) theory, although our aim is far beyond that of linking such two extremely separated energy scales. 

We have also shown that the preferred DE evolution cannot be reproduced by two perfect fluids with constant EoS parameters plus a cosmological constant, most probably because the component mimicking SQ cannot exhibit in this case thawing behavior, and the one mimicking NQ does not freeze at low redshifts.

It will be important to further explore the implications of the SQ+NQ model. We have verified that the region of parameter space leading to scalar field oscillations at low $z$ also yields very competitive results. We plan to study this region in more detail and provide a comprehensive analysis of volume effects in an extended paper.

Additionally, it would be interesting to compare our model with others that produce similar phenomenology \cite{Ye:2024ywg,Ye:2024zpk,Wolf:2025jed,Chakraborty:2025syu,Odintsov:2024woi,Nojiri:2025low,Brax:2025ahm}, particularly at the perturbative level. For instance, modifications to the Poisson equation in non-minimally coupled scalar field scenarios could lead to distinctive signatures in the growth of large-scale structures, even when the background evolution is nearly identical. We also leave this investigation for future work.

\section*{Acknowledgements}
The authors are funded by “la Caixa” Foundation (ID 100010434) and the European Union's Horizon 2020 research and innovation programme under the Marie Sklodowska-Curie grant agreement No 847648, with fellowship code LCF/BQ/PI23/11970027. They acknowledge the participation in the COST Action CA21136 “Addressing observational tensions in cosmology with systematics and fundamental physics” (CosmoVerse), and thank Prof. Joan Solà and Prof. Jaume Garriga for helpful discussions.  

\bibliographystyle{elsarticle-num}
\bibliography{biblio}

\end{document}